\documentclass[sigconf,screen,nonacm]{acmart}

\usepackage{tcolorbox}  
\usepackage{multirow}  
\usepackage{subcaption}  
\usepackage{xcolor}  
\usepackage{threeparttable}  
\usepackage{comment}  
\usepackage{microtype}  
\usepackage{colortbl}  
\definecolor{best}{HTML}{F6B26B}    
\definecolor{second}{HTML}{FCE5CD}  
\definecolor{grey}{HTML}{D1D3D4}  

\usepackage[ruled,linesnumbered]{algorithm2e}  

\newcommand{\win}[1]{\cellcolor{grey}#1}

\newcommand{\tightcolorbox}[2]{%
	{\setlength{\fboxsep}{1pt}\colorbox{#1}{#2}}%
}

\usepackage{booktabs}

\definecolor{headergray}{gray}{0.90} 
\definecolor{rowgray}{gray}{0.95}    

\AtBeginDocument{%
  }

\setcopyright{cc}
\setcctype{by}
\acmDOI{10.1145/3832783.3834351}
\acmYear{2026}
\copyrightyear{2026}
\acmConference[ASE '26]{Proceedings of the 41st IEEE/ACM International Conference on Automated Software Engineering}{October 12--16, 2026}{Munich, Germany}
\acmBooktitle{Proceedings of the 41st IEEE/ACM International Conference on Automated Software Engineering (ASE '26), October 12--16, 2026, Munich, Germany}

\begin{document}

\title{When Ambiguity Meets Atypicality: Dual-Perspective Test Input Prioritization for DNNs}
\titlenote{\textcopyright\ 2026 Haoran Li et al.
This is the author's version of \emph{When Ambiguity Meets Atypicality: Dual-Perspective Test Input Prioritization for DNNs}, accepted for publication in the Proceedings of the 41st IEEE/ACM International Conference on Automated Software Engineering (ASE '26).
ACM publication DOI: \url{https://doi.org/10.1145/3832783.3834351}.}

\author{Haoran Li}
\authornote{For correspondence, please contact Haoran Li.}
\orcid{0009-0000-8290-8167}
\affiliation{%
	\institution{Beihang University}
	\city{Beijing}
	\country{China}
}
\email{li\_haoran@buaa.edu.cn}

\author{Shihai Wang}
\correspondingauthor
\orcid{0000-0002-0241-0009}
\affiliation{%
  \institution{Beihang University}
  \city{Beijing}
  \country{China}
}
\email{wangshihai@buaa.edu.cn}

\author{Bin Liu}
\orcid{0009-0007-1991-360X}
\affiliation{%
	\institution{Beihang University}
	\city{Beijing}
	\country{China}
}
\email{liubin@buaa.edu.cn}

\author{Jialuo Chen}
\orcid{0000-0003-4322-4285}
\affiliation{%
	\institution{Zhejiang University}
	\city{Hangzhou}
	\country{China}
}
\email{chenjialuo@zju.edu.cn}

\author{Wenjing Zhu}
\orcid{0009-0008-1601-4251}
\affiliation{%
	\institution{Beihang University}
	\city{Beijing}
	\country{China}
}
\email{wenjing0616@buaa.edu.cn}

\author{Yu Liu}
\orcid{0009-0002-4457-114X}
\affiliation{%
	\institution{Beihang University}
	\city{Beijing}
	\country{China}
}
\email{liuyu415@buaa.edu.cn}

\author{Tengfei Shi}
\orcid{0009-0002-6203-2713}
\affiliation{%
	\institution{Beihang University}
	\city{Beijing}
	\country{China}
}
\email{shitfing@buaa.edu.cn}

\author{Shudi Guo}
\orcid{0009-0001-4919-0648}
\affiliation{%
	\institution{Beihang University}
	\city{Beijing}
	\country{China}
}
\email{shudiguo@buaa.edu.cn}

\renewcommand{\shortauthors}{Haoran Li, Shihai Wang, Bin Liu, Jialuo Chen, Wenjing Zhu, Yu Liu, Tengfei Shi, and Shudi Guo}

\begin{abstract}
	While Deep Neural Networks (DNNs) have achieved remarkable progress in cutting-edge domains, their inherent brittleness has become a growing concern. 
	To ensure the reliability and safety of DNN-enabled software, DNN testing has emerged as an indispensable practice. 
	Within this context, test input prioritization is essential for early fault detection and reducing labeling costs. 
	However, it remains challenging to accurately identify failure-inducing inputs. 
	Although decision ambiguity and distributional atypicality are two widely adopted perspectives for characterizing inter-class competition and intra-class typicality respectively, relying on either perspective in isolation inevitably introduces blind spots. 
	
	In this paper, we propose \textsc{DuFP} (\textbf{Du}al perspective \textbf{F}eature space \textbf{P}rioritization), a KNN density-based test input prioritization approach for DNNs that jointly incorporates both inter-class and intra-class perspectives.
	The prioritization framework of \textsc{DuFP} is built upon class-conditional density estimation. 
	Based on the estimation results, prediction correctness is characterized by an ambiguity score and an atypicality score, with the former reflecting decision ambiguity and the latter quantifying distributional atypicality. 
	A hybrid uncertainty score is then constructed by integrating both scores to guide the final prioritization. 
	We evaluate \textsc{DuFP} on prioritization and selection tasks across image and text datasets under clean, corrupted, and adversarial scenarios. 
	Experimental results demonstrate that \textsc{DuFP} effectively and efficiently prioritizes fault-inducing inputs and outperforms state-of-the-art approaches. 
\end{abstract}





\maketitle

\section{Introduction}
\label{sec_intro}
Deep Neural Networks (DNNs) have achieved significant breakthroughs in computer vision \cite{resnet}, natural language processing \cite{transformer}, and other cutting-edge domains \cite{deepspeech2, facenet, machine_translation}, becoming a milestone in artificial intelligence. 
However, the inherent brittleness of DNNs has raised growing concerns \cite{deepxplore, deepgauge}. 
Similar to traditional software, DNN-enabled systems are also susceptible to erroneous behaviors \cite{deepxplore, deepgauge}, which limits their deployment in safety-critical domains. 
Driven by these challenges, ensuring the reliability and safety of DNNs has attracted increasing research attention.

DNN testing has emerged as a prominent direction among existing efforts to ensure DNN reliability \cite{ml_testing_survey}. 
Effective and comprehensive testing relies on high-quality test data. 
Both real-world and synthetic test inputs are typically unlabeled, requiring manual labeling of their ground truth. 
However, manual labeling is costly and time-consuming. 
To ensure accuracy, labels are typically cross-validated by multiple annotators. 
In certain application domains such as medical diagnosis \cite{diabetic_retinopathy} and protein structure prediction \cite{protein_pred}, labeling further demands domain-specific knowledge. 
Furthermore, distribution shifts in test data can severely undermine the effectiveness of testing. 
To achieve thorough testing under limited time and labeling resources, test input prioritization techniques have been widely adopted so as to identify fault-inducing inputs (i.e., misclassified inputs) at early testing stages.

In recent years, diverse test prioritization techniques for DNNs have been developed. 
Existing techniques can be classified into four categories according to their underlying strategies. 
Inspired by traditional software testing, coverage-based methods \cite{deepxplore, deepgauge} employ a set of coverage metrics to measure how thoroughly test inputs activate the internal states of neural networks. 
Test suites achieving higher coverage are generally considered to provide more thorough testing. 
In addition, Surprise Adequacy (SA) was proposed to quantify the deviation of a test input from the training distribution and was subsequently incorporated into surprise-based methods \cite{sa_icse,simple_tip} for guiding prioritization. 
However, prior studies demonstrate that such metrics lack correlation with fault detection \cite{harel2020neuron, li2019structural}, suggesting that they are ineffective in guiding the identification of fault-inducing instances. 
In contrast, uncertainty metrics provide a relatively accurate quantification of prediction confidence based on output probabilities, enabling uncertainty based methods \cite{deepgini, maxp, margin} to exhibit superior effectiveness.
Nevertheless, several studies indicate that the prediction confidence of DNNs can be unreliable \cite{calibration}, motivating the development of uncertainty calibration methods FAST \cite{fast} and NNS \cite{nns}.
Despite these efforts, low-dimensional probability vectors remain relatively uninformative for characterizing the misclassification tendency of test inputs.
To address this limitation, failure mechanism-based methods \cite{ats, rts, certpri, tdpr} have been developed to analyze failure-inducing behaviors from multiple perspectives including the input domain and the feature domain.
Representative approaches include ATS \cite{ats} which considers predicted failure patterns, CertPri \cite{certpri} which focuses on adversarial robustness, RTS \cite{rts} which eliminates noisy samples, and DATIS \cite{datis} which incorporates feature neighborhood support.


Despite the advancements achieved in prioritization, existing techniques still suffer from the following limitations.
(1) \textbf{Existing prioritization approaches provide an incomplete characterization of misclassifications.}
Existing approaches predominantly assess the likelihood of misclassification from a single perspective, either decision ambiguity or distributional atypicality.
However, this inevitably subjects the prioritization to the inherent limitations of the adopted perspective.
For instance, methods such as Margin \cite{margin} and DATIS \cite{datis} are primarily designed to capture the prediction confusion between the predicted class and its competing classes.
This strategy is well suited for detecting misclassifications near the decision boundary but may overlook distributionally atypical samples with confident predictions.
Conversely, MaxP \cite{maxp} and CertPri \cite{certpri} target predictions that deviate from the typical patterns of the training distribution, but may be less effective at detecting misclassifications caused by inter-class ambiguity.
Therefore, characterizing fault-inducing inputs from either ambiguity or atypicality alone inevitably introduces blind spots.
(2) \textbf{The informative signals from feature neighborhoods are underutilized.}
The key to effective prioritization lies in accurately estimating the likelihood that the predicted class does not match the ground truth.
In practice, different classes exhibit distinct feature distributions, and the DNNs under test show varying abilities in modeling them.
For example, a single class may contain multiple modes, and class imbalance may render minority classes underrepresented.
These complexities make the overall distributional characteristics difficult to capture.
The local consistency assumption \cite{local_consistency} offers a promising direction, as the local feature neighborhood of a sample can be used to characterize its similarity to each class and even estimate class-conditional probabilities.
However, current approaches typically rely on coarse-grained distance-based metrics and lack class-wise probabilistic modeling of the feature neighborhood.

To address these limitations, we propose \textsc{DuFP}, a \textbf{Du}al perspective \textbf{F}eature space \textbf{P}rioritization method with local probabilistic modeling.
It characterizes prediction correctness from the perspectives of both decision ambiguity and distributional atypicality.
The key component of \textsc{DuFP} is the class-level hybrid uncertainty measured in the latent feature space, built upon class-conditional density estimation that embeds reliable and informative supervisory signals. 
For each test input, the hybrid uncertainty consists of an ambiguity score and an atypicality score. 
The ambiguity score quantifies the degree of confusion between the predicted class and its strongest competitor via Bayesian posterior ratios.
In contrast, the atypicality score measures how well the training distribution of the predicted class supports the given prediction via class-conditional density.
Given their complementary nature, the hybrid uncertainty is constructed by integrating both perspectives to enable more robust prioritization.
Based on this, the prioritization of the entire test set can be performed.

We extensively evaluate \textsc{DuFP} over six representative DNN benchmarks across image and text datasets under both clean and distribution shift scenarios (corrupted and adversarial). 
In comparison with 23 baselines, \textsc{DuFP} achieves the best prioritization performance in 14 of 18 evaluation cases and the second-best in 4 cases. 
On average, \textsc{DuFP} demonstrates an APFD improvement of 4.6\%--32.3\% over state-of-the-art techniques across both uncertainty based and mechanism based categories.
The evaluation results further confirm that \textsc{DuFP} exhibits high efficiency and generalizes effectively to selection tasks.

To summarize, the key contributions of this paper are as follows:
\begin{sloppypar}
\begin{enumerate}
	\item We propose \textsc{DuFP}, an effective test input prioritization method based on dual-perspective uncertainty that captures both intra-class characteristics and inter-class distinctions for the identification of fault-inducing instances.
	
	\item We extensively evaluate \textsc{DuFP} on six representative benchmarks across image and text datasets. The results demonstrate that \textsc{DuFP} consistently achieves superior performance in prioritizing test inputs under both clean and distribution shift conditions. 
	
	\item We release the implementation code and related resources to support future research, publicly available at \cite{DuFP_artifacts}.
	
\end{enumerate}
\end{sloppypar}

\section{Preliminary}
\label{sec_pre}
\subsection{Deep Neural Network}
In general, a DNN model can be regarded as a mapping function $\mathcal{F}:\mathcal{X} \to \mathcal{Y}$, 
where $\mathcal{X}$ and $\mathcal{Y}$ denote the input and output spaces, respectively. 
For the classification tasks considered in this paper, 
the output space is defined as a finite label set $\{1, 2, \cdots, C\}$. 
Given an input $x \in \mathcal{X}$, a probability vector $\mathbf{p} = \left[ p^1, p^2, \cdots, p^C \right]$ is produced by the DNN, where $p^i$ denotes the predicted probability that $x$ belongs to class $i$.
The predicted label is then determined as  $\hat{y} = \arg\max_i \{\mathbf{p}\}$, corresponding to the class with the highest predicted probability.

\subsection{Test Input Prioritization}
To improve efficiency in traditional software testing, test case prioritization (TCP) \cite{wong1997study} has been adopted for decades. 
Similarly, to address the challenges of DNN testing, test input prioritization (TIP) techniques have been introduced to identify fault-inducing inputs under limited time and labeling budgets. 
Following the formulation in traditional software testing \cite{apfd}, a formal definition of DNN test input prioritization is provided.

\textbf{\textit{Definition 1}. DNN Test Input Prioritization.} Given a test set $\mathcal{D}_T$, let \(\mathcal{S}_T\) be the set of all its possible permutations. Let \( f: \mathcal{S}_T \to \mathbb{R} \) be a scoring function that quantifies the value of each test sequence. The objective is to identify an optimal sequence \( \mathbf{s}^* \in \mathcal{S}_T \) that maximizes \( f(\mathbf{s}) \). This objective is formally defined in \eqref{eq_definition_tip}.
\begin{equation} \label{eq_definition_tip}
	\mathbf{s}^* = \arg\max_{\mathbf{s} \in \mathcal{S}_T} f(\mathbf{s})
\end{equation}
The scoring function $f$ evaluates the effectiveness of a test sequence in correctly prioritizing fault-inducing inputs, which is commonly measured by the APFD metric (detailed in Section \ref{sec_experiment_metrics}).

The technique most closely related to TIP is test input selection (TIS), which also aims to enhance DNN testing efficiency. 
The key difference is that TIS selects a subset of inputs under a limited labeling budget rather than prioritizing the entire set. 
Compared with prioritization, selection may lead to incomplete testing because some ``low-value'' test inputs are discarded. 
Nonetheless, prioritization can also be adapted to selection by choosing the top-ranked inputs in the ordering as the selection outcome. 
Therefore, this paper focuses on prioritization techniques, which are generally more applicable across testing scenarios.

\subsection{KNN Density Estimation}
Estimating the probability density of data points is fundamental to understanding the local structure of a sample space. 
K-Nearest Neighbors (KNN) density estimation \cite{knn_density} estimates the local density at a query point based on its neighboring samples.
Given a query point $x \in \mathbb{R}^d$, let $r_k(x)$ denote the distance from $x$ to its $k$-th nearest neighbor in the dataset. The probability density function (PDF) at $x$ can then be estimated as:
\begin{equation}\label{eq_knn_density}
	\hat{f}(x) = \frac{k}{n\,V\!\bigl(r_k(x)\bigr)}
\end{equation}
where $n$ is the total number of samples, $V(r) = c_d \cdot r^d$ denotes the volume of a $d$-dimensional ball with radius $r$, and $c_d$ is the corresponding volume constant.

KNN density estimation possesses two desirable properties.
First, it is non-parametric and assumes no specific form of the underlying data distribution, making it adaptable to diverse distributional characteristics.
Second, the radius $r_k(x)$ naturally contracts in dense regions and expands in sparse ones, thereby capturing local density variations without manual tuning.
These properties make KNN density estimation particularly well suited for providing informative neighborhood signals to facilitate the identification of misclassified samples.

\section{Methodology}
\label{sec_method}
\subsection{Motivation}

\begin{figure}[t]
	\centering
	
	\begin{subfigure}[t]{0.48\linewidth}
		\centering
		\includegraphics[width=\linewidth, clip, trim=0 0 30 0]{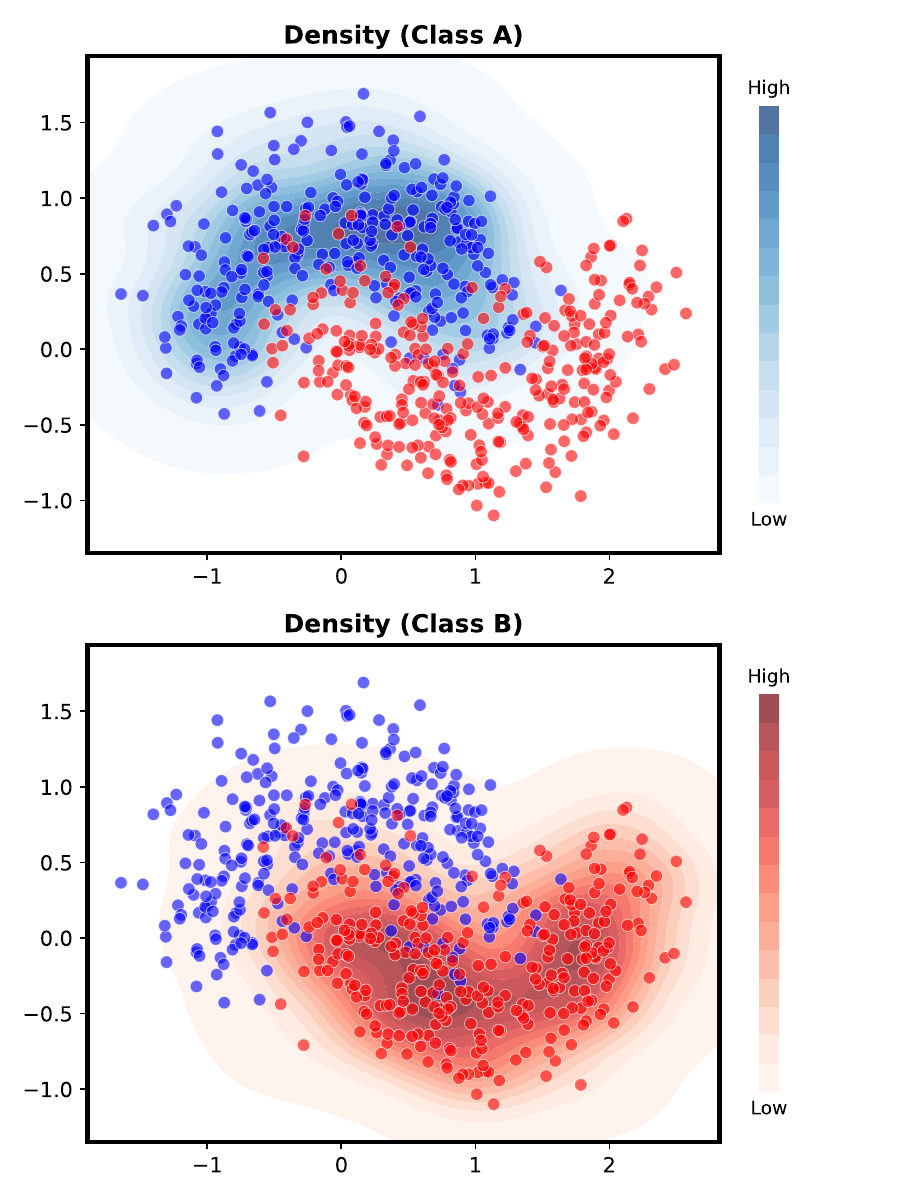}
		\caption{Class Density}
		\label{fig_toy_a}
	\end{subfigure}
	\hfill
	\begin{subfigure}[t]{0.48\linewidth}
		\centering
		\includegraphics[width=\linewidth, clip, trim=0 0 30 0]{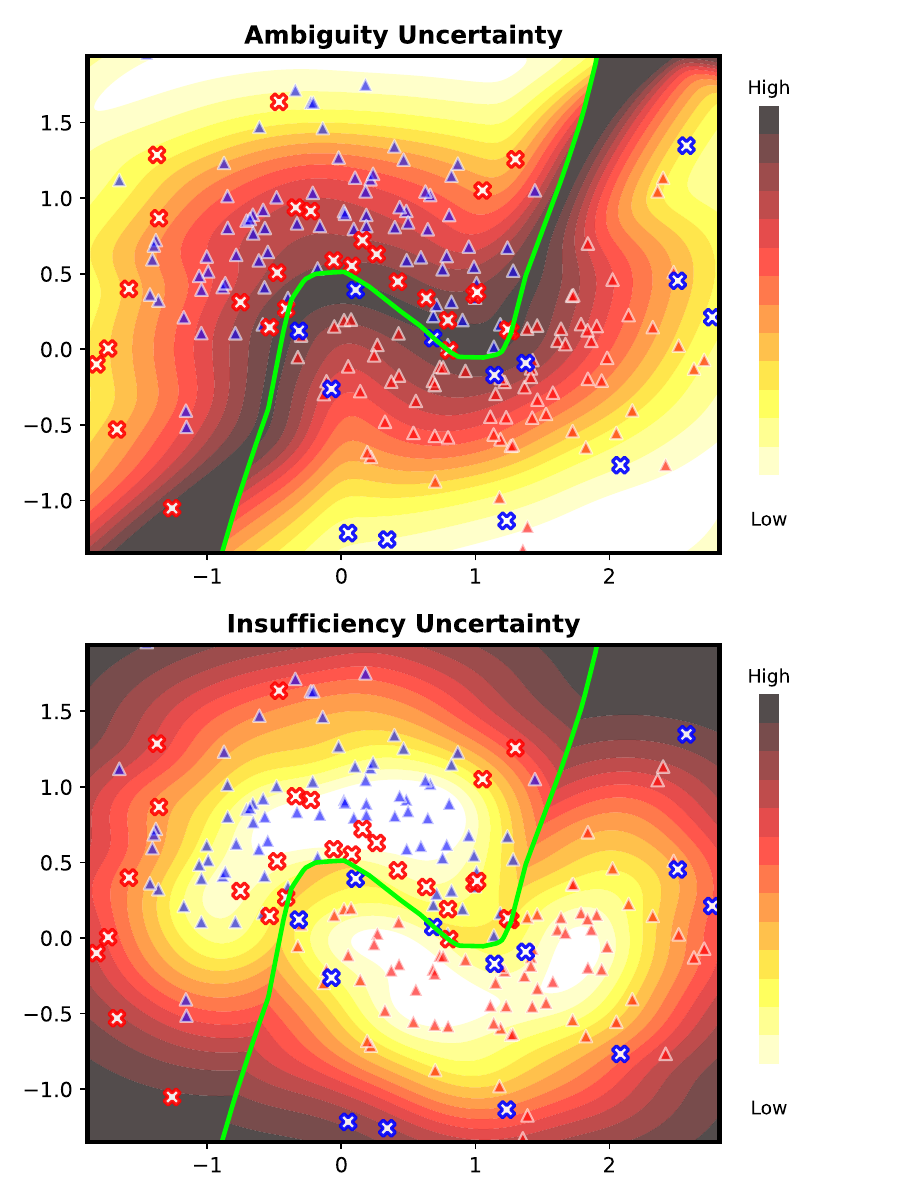}
		\caption{Single Uncertainty}
		\label{fig_toy_b}
	\end{subfigure}
		\hfill
		
	\begin{subfigure}[t]{0.75\linewidth}
		\centering
		\includegraphics[width=\linewidth, clip, trim=0 0 30 0]{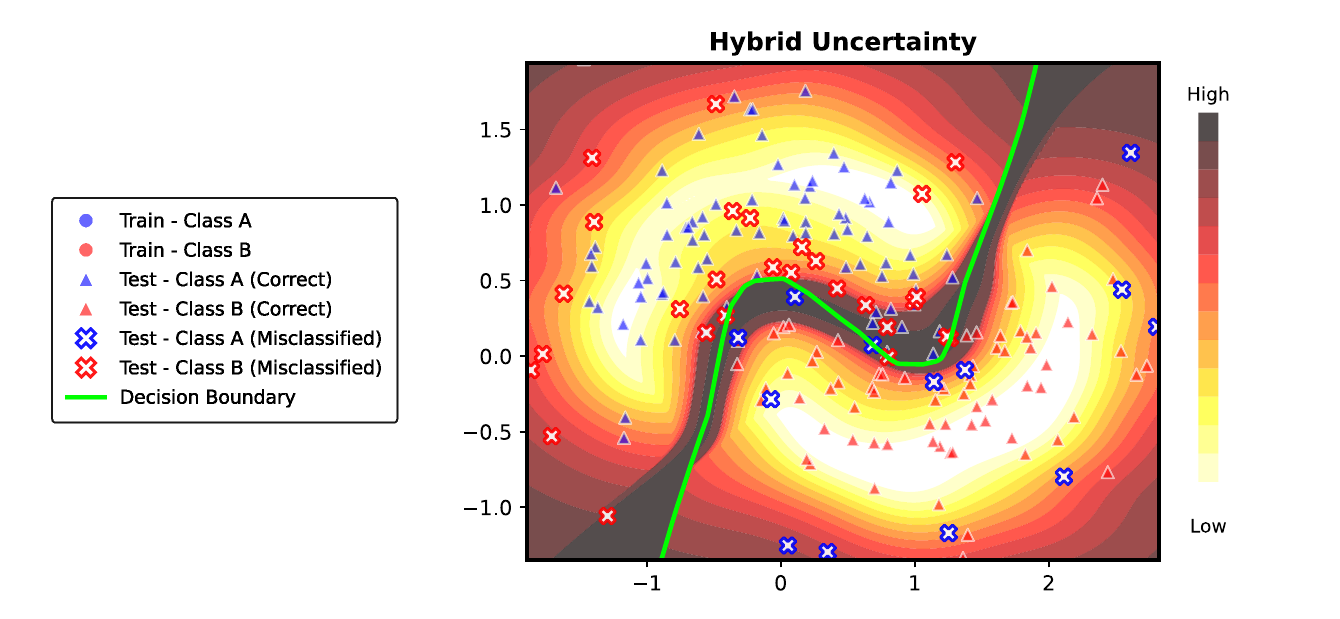}
		\caption{Hybrid Uncertainty}
		\label{fig_toy_c}
	\end{subfigure}
	\caption{The Illustration of Our Motivation Example on Two Moons Dataset.}
	\label{fig_toy}
	\Description{None.}
\end{figure}


The proposed prioritization approach is based on the following motivations. 

(1) \textbf{Uncertainty measures from dual perspectives can more robustly capture the characteristics of fault-inducing inputs.} 
Existing prioritization approaches typically measure the likelihood of misclassification from two representative perspectives: decision ambiguity and distributional atypicality.
However, relying on either perspective alone has inherent limitations.
Decision ambiguity characterizes the degree of confusion between the predicted class and its competing classes. 
It effectively identifies errors near the decision boundary but fails to detect samples that are confidently predicted yet weakly supported by the training distribution.
Distributional atypicality measures how much a given prediction deviates from the typical patterns of the training distribution. 
It effectively identifies misclassifications that deviate from the original data distribution, such as outliers and distribution-shifted samples, but fails to capture errors near the decision boundary.
Given their complementary nature, jointly incorporating both perspectives enables more robust identification of fault-inducing inputs.

To shed light on our motivation, we conduct an example experiment on the two moons dataset.
Figure~\ref{fig_toy_a} illustrates the manifold distribution of two-class training data (blue and red) and the corresponding density.
Figures~\ref{fig_toy_b} and~\ref{fig_toy_c} depict the uncertainty distributions of single-perspective and hybrid measures, respectively, overlaid with test inputs composed of both in-distribution and distribution-shifted samples (misclassified samples are highlighted with x-shaped markers).
As observed in the top panel of Figure~\ref{fig_toy_b}, the decision ambiguity perspective assigns reasonably high uncertainty to misclassifications near the decision boundary (green line) but incorrectly assigns low uncertainty to regions far from the boundary or even outside the training distribution.
Conversely, the bottom panel shows that the distributional atypicality perspective yields high uncertainty for misclassifications far from the original manifold distribution but underestimates the uncertainty near the decision boundary.
Therefore, a desirable uncertainty measure is expected to capture misclassifications near the decision boundary while assigning reasonable uncertainty to distributionally atypical samples outside the manifold.
Figure~\ref{fig_toy_c} presents the hybrid perspective that integrates both. 
The high certainty regions (light yellow) align well with the original training distributions, while the areas between and beyond the two class distributions are correctly assigned high uncertainty. 
This demonstrates a more robust misclassification detection capability.

(2) \textbf{The feature neighborhood of test inputs provides informative supervisory signals for identifying misclassifications.}
Even with calibration, uncertainty estimation that relies on low-dimensional output probabilities is regarded as unreliable. 
This limitation stems from the difficulty of identifying failure mechanisms solely based on model outputs. 
In other words, the causes cannot be traced from the results. 
In white-box testing scenarios, both model information and the supervision from training data are accessible. 
The feature embeddings extracted by the model with ground-truth can be regarded as a natural combination of both.
Under the local consistency assumption \cite{local_consistency}, a test input's feature embedding tends to lie close to samples of the same class, which provides informative evidence for identifying misclassifications.
Indeed, recent studies leveraging coarse-grained neighborhood information~\cite{datis, nns} demonstrate superior performance.
Therefore, probabilistic modeling of the class-wise distribution within the feature neighborhood is expected to facilitate a more fine grained assessment of misclassification tendency.

\subsection{Overview}
\begin{figure*}[t]
	\centering
	\includegraphics[width=0.9\linewidth, clip, trim=10 10 10 10]{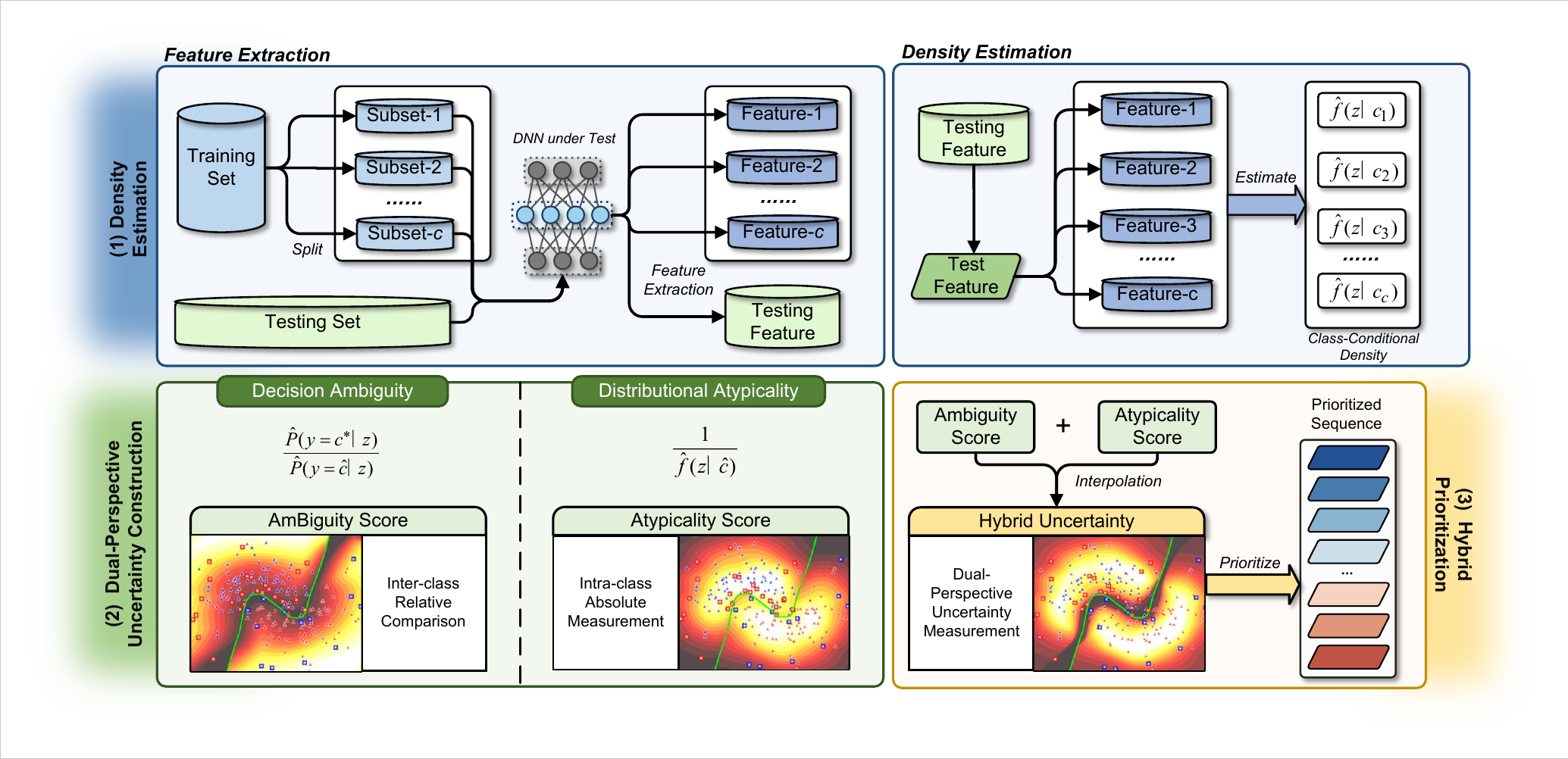}
	\caption{The Workflow of \textsc{DuFP}.}
	\label{fig_workflow}
	\Description{None.}
\end{figure*}

Building on the above motivations, we propose \textsc{DuFP}, a prioritization method for DNNs based on dual-perspective uncertainty.
As illustrated in Figure \ref{fig_workflow}, the workflow of \textsc{DuFP} consists of three primary stages.

\textbf{(1) Class-Conditional Density Estimation}. 
To provide feature neighborhood information for subsequent uncertainty characterization, the feature representations of training samples are first extracted and grouped by class.
Class-wise KNN density estimators are then constructed based on these features to estimate the class-conditional density of each test input.

\textbf{(2) Uncertainty Quantification from Two Perspectives}. 
The prediction uncertainty is then quantified through the ambiguity score and the atypicality score, which characterize decision ambiguity and distributional atypicality, respectively.
The ambiguity score is derived from Bayesian posterior probabilities, capturing the degree of confusion of the test feature between the predicted class and its strongest competitor.
The atypicality score is derived from the class-conditional density, measuring how well the training distribution of the predicted class supports the given prediction.

\textbf{(3) Prioritization with Hybrid Uncertainty}. 
Finally, a hybrid uncertainty is constructed by integrating both perspectives, jointly accounting for inter-class relative comparison and intra-class absolute measurement. 
The entire test set is subsequently prioritized based on this hybrid uncertainty.

\subsection{Class-Conditional Density Estimation}
\label{sec_method_density}
To characterize fine-grained distributional information embedded in the feature neighborhood of test inputs, the class-conditional density is estimated for each test input.
This requires the feature representations of the training set to be extracted in advance.

\textbf{Feature Extraction.} 
Following common practice in prior work \cite{frosst2019analyzing, trust_score}, the representation space of the penultimate layer is adopted as the feature space in this study.
For each instance $x$, the penultimate-layer activation is flattened and $\ell_2$-normalized to obtain the feature representation $z$.
The entire training set $\mathcal{D}$ is then partitioned into $C$ class-specific subsets $\{\mathcal{D}_c\}_{c=1}^{C}$ according to ground-truth labels, and the corresponding feature representations are precomputed as $\mathcal{Z}_c$ for subsequent density estimation.

\textbf{Density Estimation.} 
With these precomputed feature sets, the class-conditional density of a given test feature $z$ with respect to class $c$ can be estimated.
For downstream uncertainty quantification to be reliable, density estimates should be comparable across classes.
However, due to potential class imbalance in the training set, using a fixed $k$ across classes may lead to incomparable density estimates.
To ensure a comparable neighborhood scale across classes with varying sample sizes, the neighbor count $k_c$ is set proportionally to the class size $n_c$, i.e., $k_c = n_c \cdot \eta$, where $\eta \in (0, 1]$ denotes the neighbor ratio relative to the class size.
Under this setting, the class-conditional density is formulated as:
\begin{equation}\label{eq_density}
	\hat{f}(z \mid c) = \frac{\eta}{V_d\, r_c(z)^d}
\end{equation}
where $r_c(z)$ denotes the distance from $z$ to its $k_c$-th nearest neighbor in $\mathcal{Z}_c$. 
This formulation eliminates the explicit class-frequency factor, ensuring that density values across classes reflect purely geometric proximity rather than class prevalence.

\subsection{Uncertainty from Two Perspectives}
To comprehensively characterize the misclassification tendency from complementary perspectives, \textsc{DuFP} quantifies prediction uncertainty through both decision ambiguity and distributional atypicality.

\textbf{Ambiguity Score.}
The key to estimating decision ambiguity lies in how distinguishable the predicted class is from its competing classes within the feature neighborhood of a test input.
Intuitively, a prediction is more likely to be incorrect when the test feature exhibits a higher probability of belonging to other classes than to the predicted class.
Built on the class-conditional densities defined in \eqref{eq_density}, the posterior probability that feature $z$ belongs to class $c$ is derived as:
\begin{equation}\label{eq_posterior}
\hat{P}(y = c \mid z) = \frac{\hat{f}(z \mid c)\,\pi_c}{\sum_{c'} \hat{f}(z \mid c')\,\pi_{c'}}
\end{equation}
where $\pi_c = n_c / N$ denotes the class prior estimated from the training set.

The likelihood ratio has been demonstrated to serve as a Neyman-Pearson optimal metric for OOD detection \cite{flats}, which is essentially a binary hypothesis testing problem that only distinguishes between in-distribution and out-of-distribution samples.
For fault-inducing input identification, the decision process is considerably more complex, as the ground truth of a misclassified sample can be any class other than the predicted one.

To simplify this multi-class problem, we follow the top-2 concentration assumption \cite{when_to_abstain}, which states that predictions of modern high-performance DNNs are typically concentrated on the top two classes.
Under this assumption, the misclassification probability can be approximated as $\hat{P}(y \neq \hat{c} \mid z) \approx \hat{P}(y = c^* \mid z)$, where $\hat{c}$ denotes the predicted class and $c^* = \arg\max_{c \neq \hat{c}} \hat{P}(y=c \mid z)$ is the strongest competing class.
Therefore, decision ambiguity can be effectively measured by evaluating the prediction confidence between the two most probable classes.
Accordingly, the ambiguity score is defined as:
\begin{equation}\label{eq_ambiguity}
	s_{\mathrm{amb}}(z) 
	= \frac {\hat P(y=c^* \mid z)} { \hat P(y=\hat c \mid z)}
\end{equation}
A higher $s_{\mathrm{amb}}$ indicates that the competing class's posterior approaches or exceeds that of the predicted class, suggesting that the test feature lies in a high-risk region near the decision boundary.

\textbf{Atypicality Score.}
The ambiguity score captures inter-class competition but is insensitive to cases where the predicted class dominates its competitors while the test input lies in a low-density region of that class.
To address this blind spot, we introduce a complementary metric that measures how well the predicted class distribution supports the given test input.
Unlike the ambiguity score, which performs a relative comparison between classes, the atypicality score provides an absolute assessment within the predicted class alone.
Specifically, the atypicality score is defined as the reciprocal of the class-conditional density under the predicted class:
\begin{equation}\label{eq_atypicality}
	s_{\mathrm{atyp}}(z)  =  \frac{1}{\hat f(z \mid \hat c)}
\end{equation}
A higher $s_{\mathrm{atyp}}$ indicates that the test feature resides in a low-density region of the predicted class distribution, suggesting that the prediction lacks sufficient support from the training distribution and is more likely to be a misclassification.


\subsection{Prioritization with Hybrid Uncertainty}
The ambiguity and atypicality scores essentially capture inter-class relative comparison and intra-class absolute measurement, respectively.
To take advantage of both perspectives, \textsc{DuFP} integrates the two scores into a hybrid uncertainty to guide the overall prioritization.
Due to the scale difference between the two scores, both are log-transformed and then independently min-max normalized to $[0,1]$ over the entire test set before integration.
Taking the ambiguity score as an example, let $a_i = \log s_{\mathrm{amb}}(z_i)$, and the normalized component is computed as:
\begin{equation} \label{eq_normalize}
	\tilde{s}_{\mathrm{amb}}(z_i) = \frac{a_i - \min_{j} a_j}{\max_{j} a_j - \min_{j} a_j + \epsilon}
\end{equation}
where $j$ ranges over all test inputs, and $\epsilon$ is a small constant guarding against a zero denominator.
The atypicality component $\tilde{s}_{\mathrm{atyp}}$ is computed analogously.
A linear interpolation is then adopted for integration, and the hybrid uncertainty score for a given test feature $z$ is defined as:
\begin{equation} \label{eq_hybrid}
	{s}_{\mathrm{hy}}(z) = (1 - \lambda) \tilde{s}_{\mathrm{amb}} + \lambda \tilde{s}_{\mathrm{atyp}}
\end{equation}
The balance parameter $\lambda \in [0, 1]$ controls the trade-off between the two perspectives.
A larger $\lambda$ makes the hybrid uncertainty more biased toward the atypicality score, whereas a smaller $\lambda$ emphasizes the ambiguity score. 
The sensitivity of \textsc{DuFP} to this parameter is further analyzed in RQ4.
The entire test set is then sorted in descending order of the hybrid uncertainty score, so that inputs more likely to be misclassified are prioritized for earlier testing.

\section{Experiments Design}
\label{sec_experiment}

\subsection{Datasets and Models}
We conducted a comprehensive evaluation across four image datasets and two text datasets. 
Among the image datasets, Fashion-MNIST \cite{fashionmnist} contains ten clothing categories with 60,000 training and 10,000 test grayscale images of size $28 \times 28$.
CIFAR-10 \cite{cifar10} is a widely used image dataset containing ten categories of vehicles and animals, with each sample represented as a $32 \times 32$ RGB image.
SVHN \cite{svhn} consists of house number images spanning ten digit classes from 0 to 9, with 73,257 training samples and 26,032 test samples in the same image format as CIFAR-10.
For these image datasets, representative DNN architectures were adopted from official repositories released in previous studies \cite{ats,rts,nss}, including VGG-16 \cite{vgg} and ResNet-20 \cite{resnet}.
Additionally, ImageNet-100 \cite{imagenet} was used to evaluate the effectiveness of the proposed method in more complex multi-class classification tasks.
ImageNet-100 is a lightweight subset of ImageNet that contains 100 classes and preserves the original image resolution \cite{imagenet100_kaggle}.
It has been widely adopted in studies on DNN robustness and safety due to its reduced validation complexity \cite{smir, laidlaw2020perceptual, jain_2024_cvpr}.
Following \cite{deit_hf}, DeiT \cite{deit} was adopted as the representative Vision Transformer (ViT) model for this dataset.
Since the model was originally trained on the full ImageNet dataset, its classifier head was adapted to ImageNet-100 for compatibility.

To evaluate the generalizability of \textsc{DuFP} on text data, AGNews and DBPedia \cite{agnews_and_dbpedia} were adopted.
AGNews \cite{agnews_and_dbpedia} is a news classification dataset with four categories, containing 120,000 training samples and 7,600 test samples.
DBPedia \cite{agnews_and_dbpedia} is a relatively large-scale text classification dataset constructed from Wikipedia, containing 14 ontology classes with 560,000 training items and 70,000 test items.
Each item consists of a title and corresponding content.
For these two text datasets, two BERT models \cite{bert} with publicly available fine-tuned weights from HuggingFace \cite{agnews_bert, dbpedia_bert} were employed.

To ensure a fair comparison, all DNN models were directly adopted from prior studies or public repositories without additional training.
All datasets and models are summarized in Table \ref{tab_models}.

\begin{table}[ht]
	\centering
	\caption{Datasets and Models Used in Our Evaluation}
	\label{tab_models}
	\resizebox{\columnwidth}{!}{ 
		\begin{tabular}{cccccccccccccccc}
			\toprule
			\textbf{Dataset Type} & \textbf{Dataset} & \textbf{\#Classes} & \textbf{Candidate Size} & \textbf{Model} & \textbf{Test Accuracy} \\
			\midrule
			
			\multirow{4.5}{*}{\centering {Image Datasets}} 
			
			
			& \multirow{1}{*}{\centering {Fashion-MNIST \cite{fashionmnist}}} & \multirow{1}{*}{\centering {10}} 
			& 10,000 & ResNet20 \cite{resnet} & 92.47\% \\
			\cmidrule(r){2-6}
			
			& \multirow{1}{*}{\centering {CIFAR-10 \cite{cifar10}}} & \multirow{1}{*}{\centering {10}} 
			& 10,000 & ResNet20 \cite{resnet} & 80.30\% \\
			\cmidrule(r){2-6}
			
			& \multirow{1}{*}{\centering {SVHN \cite{svhn}}} & \multirow{1}{*}{\centering {10}} 
			& 26,032 & VGG16 \cite{vgg} & 95.37\% \\
			\cmidrule(r){2-6}
			
			
			& \multirow{1}{*}{\centering {ImageNet-100 \cite{imagenet100_kaggle}}} & \multirow{1}{*}{\centering {100}} 
			& 5,000 & DeiT \cite{deit} & 88.78\% \\
			
			\midrule
			
			\multirow{2}{*}{\centering {Text Datasets}} 
			
			& \multirow{1}{*}{\centering {AGNews \cite{agnews_and_dbpedia}}} & \multirow{1}{*}{\centering {4}} 
			& 7,600 & BERT \cite{bert} & 93.45\% \\
			\cmidrule(r){2-6}
			
			& \multirow{1}{*}{\centering {DBPedia \cite{agnews_and_dbpedia}}} & \multirow{1}{*}{\centering {14}} 
			& 70,000 & BERT \cite{bert} & 99.03\% \\
			
			\bottomrule
		\end{tabular}
		
	} 
\end{table}

\subsection{Candidate Set Construction}
The evaluation was conducted under clean, corrupted, and adversarial conditions.
These three conditions are designed to be complementary: the nominal candidate set represents in-distribution data under ideal conditions, while the corrupted and adversarial candidate sets jointly represent data under distribution shift. 
Specifically, corrupted data captures natural distribution shifts caused by environmental factors, and adversarial data addresses malicious distribution shifts induced by deliberate attacks.

\textbf{Nominal Candidate Set.} 
Following prior studies \cite{datis, tdpr, certpri, simple_tip}, the original test set was directly adopted as the nominal candidate set in our experiments.

\textbf{Corrupted Candidate Set.}
For image datasets, corruption operators follow the widely adopted benchmark \cite{imagenet_c}, covering multiple corruption types that span noise, blur, weather, and digital distortions. 
This benchmark has been extensively validated to reflect realistic distribution shifts encountered in operational environments.
Specifically, Fashion-C was obtained from public repositories \cite{simple_tip}, while CIFAR10-C, ImageNet-100-C, and SVHN-C were generated following the framework of \cite{imagenet_c}.
For text datasets, corruption operators were defined following the benchmark \cite{simple_tip}, including wrong autocompletions, erroneous auto-corrections, single-word level translations, and single-letter typos.
AGNews-C and DBPedia-C were constructed accordingly using the same approach.

\textbf{Adversarial Candidate Set.}
Following \cite{nns, prima}, five widely used adversarial attack methods were adopted for image datasets, including FGSM \cite{fgsm}, BIM \cite{bim}, PGD \cite{pgd}, CW \cite{cw}, and DeepFool \cite{deepfool}.
For text datasets, two classical adversarial attack methods, TextFooler \cite{textfooler} and PWWS \cite{pwws}, were employed.
Given the original test set, each sample was randomly subjected to one of the aforementioned attack types with equal probability. 
The attack procedures for image and text datasets were implemented using the torchattacks \cite{torchattacks} and textattack \cite{textattack} libraries, respectively.
Following \cite{certpri, prima}, the adversarial candidate set was constructed by randomly selecting half of the natural test inputs and half of the adversarial inputs.

\subsection{Baselines}
A comprehensive set of 23 test prioritization techniques across four categories was adopted as baselines.

\textbf{Coverage-Based Methods}. 
Structural coverage metrics have been widely adopted in prior studies \cite{deepgini} but have been shown to be less effective.
Two representative metrics were employed, including NAC \cite{deepxplore} as the first proposed coverage criterion and KMNC \cite{deepgauge} as a fine-grained variant.
In the experiments, they were evaluated under both CTM and CAM strategies \cite{deepgini}. 
Hyperparameter values followed the settings used in prior studies \cite{nss, deepgauge}.

\textbf{Surprise-Based Methods}. 
Surprise adequacy (SA) metrics were employed, including the standard LSA \cite{sa_tosem} and the per-class variants PC-LSA, PC-DSA, PC-MLSA, PC-MDSA, and PC-MMDSA \cite{simple_tip}.
Following \cite{simple_tip}, SA values were transformed into coverage metrics and prioritized under the CAM strategy.

\textbf{Uncertainty-Based Methods}. 
The classical uncertainty metrics, including MaxP \cite{maxp}, Margin \cite{margin}, Entropy \cite{deepgini}, and DeepGini \cite{deepgini}, were employed. 
In addition, two uncertainty calibration methods FAST \cite{fast} and NNS \cite{nns} were also included.

\textbf{Failure Mechanism-Based Methods}. 
Recently proposed methods that explore failure mechanisms from different perspectives were also selected, including CertPri \cite{certpri}, ATS \cite{ats}, RTS \cite{rts}, NSS \cite{nss}, and PRIMA \cite{prima}. 
Additionally, selection approaches MCP \cite{mcp} and DATIS \cite{datis} were adopted exclusively for selection tasks (RQ2), as they are unable to produce complete ranking sequences required for prioritization tasks (RQ1).
However, due to the two-stage selection workflow of DATIS that combines ranking and redundancy removal, a variant excluding the second stage was introduced and denoted as DATIS$_{\text{r}}$. 
This variant is capable of producing ranking sequences and was included in the comparison for RQ1.

In addition, several related approaches were excluded for the following reasons.
NBC, SNAC, and TKNC \cite{deepgauge} belong to the same structural coverage family as the adopted NAC and KMNC, for which prior studies have reported limited correlation with fault detection \cite{harel2020neuron, li2019structural}.
LOF \cite{lof} originates from density-based outlier detection and has not been fully explored as a prioritization technique in this field.
TDPR \cite{tdpr} was excluded because it requires learning trajectories collected during the entire training process, which is incompatible with the well\mbox{-}trained models adopted in this study.
SETS \cite{sets} targets small-budget test selection rather than complete prioritization, and its time cost increases sharply with the labeling budget.
CES \cite{ces} aims to estimate overall test accuracy through representative sampling rather than to expose faults early.
Nevertheless, the 23 adopted baselines span all four categories and sufficiently cover the state of the art in the studied setting.
For all baseline methods, hyperparameters were configured following the settings reported in previous studies.


\subsection{Evaluation Metrics}
\label{sec_experiment_metrics}
Two widely used metrics were adopted for evaluation.

\textbf{Average Percentage of Fault Detection (APFD)} \cite{apfd} is a standard metric for measuring the overall effectiveness of a prioritized test input sequence. 
Given an ordered test input set, the APFD value is calculated as shown in \eqref{eq_apfd}. 
\begin{equation}\label{eq_apfd}
	APFD = 1 - \frac{TF_1 + TF_2 + \cdots + TF_M}{NM} + \frac{1}{2N}
\end{equation}
where $N$ denotes the total number of test inputs, 
$M$ denotes the number of misclassified instances, 
and $TF_i$ represents the position at which the $i$-th fault is detected. 
The APFD value ranges from 0 to 1, with higher values indicating earlier fault detection.

\textbf{Test-Relative Coverage (TRC)} \cite{testrank} is another commonly used evaluation metric. 
In contrast to APFD, which measures the overall performance of a prioritized test suite, 
TRC evaluates fault detection under a given test budget. 
Given a test budget and the corresponding test suite, TRC is defined as: 
\begin{equation}\label{eq_trc}
	TRC = \frac{|F_S|}{\min(|D_S|, |F|)}
\end{equation}
where $|D_S|$ and $|F_S|$ denote the total number of inputs and the number of faults in the selected set respectively, and $|F|$ denotes the total number of faults in the entire candidate set.
A higher TRC value indicates that more misclassified instances are identified within the given budget.


\subsection{Research Questions}
We comprehensively evaluated \textsc{DuFP} through the following research questions:

\textbf{RQ1. Prioritization}: How effectively does \textsc{DuFP} perform in prioritizing test inputs?

\textbf{RQ2. Selection}: How effectively does \textsc{DuFP} perform in selecting test inputs?

\textbf{RQ3. Efficiency}: How efficient is \textsc{DuFP} in prioritizing test inputs?

\textbf{RQ4. Ablation Study}: How does each design element of \textsc{DuFP} contribute to its effectiveness?

\subsection{Implementation}
We conducted the experiments on an Ubuntu 20.04 server with an Intel Xeon Platinum 8458P CPU, an RTX 4090 GPU, and 120\,GB of memory. 
The implementation was performed in Python 3.8 with PyTorch 1.12.0.
To mitigate the effect of randomness, each experimental setting was repeated five times.

\section{Results}
\label{sec_results}
\subsection{RQ1: Prioritization}
To extensively evaluate the effectiveness of \textsc{DuFP} in prioritization, we conducted experiments on 18 subjects (6 benchmarks $\times$ 3 data scenarios) against 22 baselines.
The hyperparameters were set to the recommended values reported in the corresponding studies \cite{nss, deepgauge, fast, ats, rts}. 
It is worth noting that certain baseline methods could not be applied to all evaluation scenarios.
Specifically, the official implementations of RTS, PRIMA, and CertPri were not applicable to BERT architectures, and RTS and PRIMA were additionally incompatible with the ViT architecture (ImageNet-100).
Moreover, certain coverage-based methods incurred excessive execution time on large-scale models (e.g., KMNC-CAM required over 100 hours on VGG16) and proved both inefficient and ineffective. 
As this contradicts the objective of test prioritization, such configurations were excluded from the evaluation.
The APFD metric was adopted as the measure of effectiveness in this evaluation.

\begin{table*}[t]
	\centering
	\caption{APFD Results on Nominal (Nom.), Corrupted (Corr.), and Adversarial (Adv.) Data.}
	\label{tab_apfd}
	\resizebox{\textwidth}{!}{
		\begin{threeparttable}
			\setlength{\tabcolsep}{3pt}   
			\renewcommand{\arraystretch}{1.0}
			
			\begin{tabular}{c c *{18}{c} | ccc}
				\toprule
				\multicolumn{2}{c}{\multirow{2}{*}{\textbf{Method}}} &
				\multicolumn{3}{c}{Fashion} & \multicolumn{3}{c}{CIFAR\mbox{-}10} & \multicolumn{3}{c}{SVHN} & \multicolumn{3}{c}{ImageNet\mbox{-}100} & \multicolumn{3}{c}{AGNews} & \multicolumn{3}{c|}{DBPedia} & Stat. & \multicolumn{2}{c}{Avg.} \\
				\cmidrule(lr){3-5}\cmidrule(lr){6-8}\cmidrule(lr){9-11}\cmidrule(lr){12-14}\cmidrule(lr){15-17}\cmidrule(lr){18-20}\cmidrule(lr){21-21}\cmidrule(lr){22-23}
				& & Nom. & Corr. & Adv. & Nom. & Corr. & Adv. & Nom. & Corr. & Adv. & Nom. & Corr. & Adv. & Nom. & Corr. & Adv. & Nom. & Corr. & Adv. & W/T/L & Imp. & Rank \\
				\midrule
				\multirow{4}{*}{\rotatebox{90}{Coverage}}
				& NAC\mbox{-}CTM   
				& \win{}31.944 & \win{}52.213 & \win{}63.488 & \win{}46.242 & \win{}48.190 & \win{}54.893 & \win{}67.048 & \win{}69.390 & \textbf{69.396} & \win{}67.948 & \win{}66.700 & \win{}67.107 & \win{}52.285 & \win{}42.714 & \win{}42.687 & \win{}28.142 & \win{}28.843 & \win{}25.431 & 17/0/1 & 73.8\% & 13.6 \\
				
				& NAC\mbox{-}CAM   
				& \win{}49.929 & \win{}49.853 & \win{}49.904 & \win{}50.023 & \win{}49.784 & \win{}50.096 & \win{}50.930 & \win{}50.277 & \win{}49.954 & \win{}49.652 & \win{}49.967 & \win{}49.931 & \win{}49.683 & \win{}49.318 & \win{}50.338 & \win{}50.779 & \win{}50.129 & \win{}50.009 & 18/0/0 & 57.9\% & 17.3 \\
				
				
				
				
				
				& KMNC\mbox{-}CTM  
				& \win{}50.718 & \win{}45.529 & \win{}26.697 & \win{}49.701 & \win{}47.770 & \win{}33.564 & \win{}49.741 & \win{}45.694 & \win{}27.506 & \win{}52.966 & \win{}52.571 & \win{}54.079 & \win{}49.523 & \win{}47.935 & \win{}47.181 & \win{}59.728 & \win{}39.707 & \win{}44.254 & 18/0/0 & 76.5\% & 18.6 \\
				
				& KMNC\mbox{-}CAM  
				& \win{}41.030 & \win{}46.740 & \win{}54.782 & \win{}39.700 & \win{}41.264 & \win{}53.445 & -- & -- & -- & -- & -- & -- & \win{}76.925 & \win{}67.435 & \win{}62.465 & -- & -- & -- & 9/0/0 & 42.0\% & 14.9 \\
				
				
				\midrule
				\multirow{6}{*}{\rotatebox{90}{Surprise}}
				& LSA              
				& \win{}52.413 & \win{}50.653 & \win{}52.620 & \win{}49.744 & \win{}49.913 & \win{}51.180 & \win{}49.628 & \win{}50.179 & \win{}49.915 & \win{}55.658 & \win{}55.473 & \win{}56.320 & \win{}49.802 & \win{}49.804 & \win{}49.885 & -- & -- & -- & 15/0/0 & 50.1\% & 15.7 \\
				
				& PC\mbox{-}LSA    
				& \win{}55.487 & \win{}49.975 & \win{}50.061 & \win{}52.955 & \win{}52.008 & \win{}49.992 & \win{}49.628 & \win{}50.179 & \win{}49.915 & \win{}49.874 & \win{}49.930 & \win{}49.431 & \win{}49.802 & \win{}49.804 & \win{}49.885 & \win{}50.643 & \win{}49.786 & \win{}49.954 & 18/0/0 & 56.5\% & 16.7 \\
				
				& PC\mbox{-}MLSA   
				& \win{}53.859 & \win{}50.455 & \win{}52.281 & \win{}52.626 & \win{}51.996 & \win{}51.828 & \win{}53.685 & \win{}52.433 & \win{}51.276 & \win{}56.905 & \win{}56.222 & \win{}55.966 & \win{}54.152 & \win{}54.455 & \win{}51.864 & \win{}54.478 & \win{}52.438 & \win{}50.862 & 18/0/0 & 48.3\% & 12.6 \\
				
				& PC\mbox{-}DSA    
				& \win{}49.327 & \win{}49.947 & \win{}50.103 & \win{}50.266 & \win{}50.013 & \win{}50.119 & \win{}50.334 & \win{}50.331 & \win{}49.941 & \win{}50.184 & \win{}50.385 & \win{}49.655 & \win{}50.515 & \win{}50.212 & \win{}49.837 & -- & -- & -- & 15/0/0 & 54.5\% & 17.4 \\
				
				& PC\mbox{-}MDSA   
				& \win{}51.383 & \win{}50.047 & \win{}51.517 & \win{}51.461 & \win{}50.690 & \win{}51.795 & \win{}50.723 & \win{}51.539 & \win{}50.640 & \win{}55.645 & \win{}55.749 & \win{}54.591 & \win{}61.383 & \win{}56.644 & \win{}52.747 & \win{}52.969 & \win{}51.307 & \win{}50.543 & 18/0/0 & 49.6\% & 13.8 \\
				
				& PC\mbox{-}MMDSA  
				& \win{}50.333 & \win{}49.873 & \win{}51.951 & \win{}50.736 & \win{}50.444 & \win{}51.644 & \win{}50.965 & \win{}50.825 & \win{}51.191 & \win{}52.223 & \win{}52.568 & \win{}50.814 & \win{}58.791 & \win{}56.737 & \win{}53.005 & \win{}51.302 & \win{}50.923 & \win{}50.085 & 18/0/0 & 52.4\% & 14.3 \\

				\midrule
				\multirow{6}{*}{\rotatebox{90}{Uncertainty}}
				& DeepGini         
				& \win{}83.680 & \win{}54.753 & \win{}51.635 & \win{}77.158 & \win{}70.062 & \win{}49.647 & \win{}88.724 & \win{}87.760 & \win{}50.283 & \win{}86.894 & \win{}86.012 & \win{}84.342 & \win{}85.101 & \win{}73.187 & \win{}75.818 & \win{}90.671 & \win{}80.690 & \win{}82.667 & 18/0/0 & 5.9\% & 6.3 \\
				
				& MaxP             
				& \win{}83.678 & \win{}54.753 & \win{}51.633 & \win{}77.144 & \win{}70.042 & \win{}49.646 & \win{}88.716 & \win{}87.755 & \win{}50.310 & \win{}86.872 & \win{}86.064 & \win{}84.373 & \win{}85.100 & \win{}73.185 & \win{}75.819 & \win{}90.671 & \win{}80.679 & \win{}82.669 & 18/0/0 & 5.8\% & 6.7 \\
				
				& Margin           
				& \win{}83.867 & \win{}54.748 & \win{}51.730 & \win{}77.182 & \win{}70.103 & \win{}49.768 & \win{}88.764 & \win{}87.790 & \win{}50.397 & \win{}87.010 & \win{}86.344 & \win{}84.500 & \win{}85.014 & \win{}73.016 & \win{}75.766 & \win{}90.837 & \win{}80.501 & \win{}82.728 & 18/0/0 & 5.8\% & 5.8 \\
				
				& Entropy         
				& \win{}83.922 & \win{}54.754 & \win{}52.037 & \win{}77.821 & \win{}70.795 & \win{}50.796 & \win{}88.669 & \win{}87.729 & \win{}50.255 & \win{}86.222 & \win{}84.956 & \win{}83.563 & \win{}85.132 & \win{}73.236 & \win{}75.832 & \win{}90.625 & \win{}80.783 & \win{}82.641 & 18/0/0 & 5.7\% & 5.8 \\
				
				& FAST             
				& \win{}81.963 & \win{}55.045 & \win{}51.541 & \win{}76.782 & \win{}70.613 & \win{}49.300 & \win{}86.633 & \win{}85.606 & \win{}49.135 & \win{}80.638 & \win{}79.036 & \win{}77.294 & \win{}84.778 & \win{}73.552 & \win{}75.971 & \win{}90.387 & \win{}80.672 & \win{}82.305 & 18/0/0 & 7.9\% & 8.3 \\
				
				& NNS              
				& \win{}80.632 & \win{}53.877 & \win{}58.309 & \win{}75.795 & \win{}70.241 & \win{}54.417 & \win{}88.539 & \win{}87.557 & \win{}53.420 & \win{}86.977 & \win{}86.248 & \win{}84.560 & \win{}84.613 & \win{}72.527 & \win{}75.732 & \win{}89.998 & \win{}80.281 & \win{}82.422 & 18/0/0 & 4.6\% & 6.3 \\
				
				\midrule
				
				\multirow{7}{*}{\rotatebox{90}{Mechanism}}
				& NSS              
				& \win{}57.026 & \win{}51.961 & \win{}59.214 & \win{}53.434 & \win{}53.486 & \win{}54.407 & \win{}54.089 & \win{}55.290 & \win{}56.643 & \win{}67.366 & \win{}67.412 & \win{}69.057 & -- & -- & -- & -- & -- & -- & 12/0/0 & 32.3\% & 10.7 \\
				
				& ATS              
				& \win{}69.282 & \win{}53.583 & \win{}56.327 & \win{}67.073 & \win{}62.795 & \win{}54.176 & \win{}81.997 & \win{}81.414 & \win{}58.088 & \win{}82.623 & \win{}81.133 & \win{}79.888 & \win{}65.181 & \win{}58.942 & \win{}67.778 & \win{}72.795 & \win{}69.847 & \win{}75.050 & 18/0/0 & 15.1\% & 9.2 \\
				
				& RTS              
				& \win{}77.565 & \textbf{58.897} & \win{}61.088 & \win{}70.658 & \win{}65.144 & \win{}50.315 & \win{}79.881 & \win{}78.903 & \win{}45.908 & -- & -- & -- & -- & -- & -- & -- & -- & -- & 8/0/1 & 13.5\% & 10.6 \\
				
				& CertPri          
				& \win{}82.198 & \win{}53.238 & \win{}48.769 & \win{}72.709 & \win{}65.595 & \win{}48.951 & \win{}88.930 & \win{}87.879 & \win{}52.219 & -- & -- & -- & -- & -- & -- & -- & -- & -- & 9/0/0 & 12.4\% & 10.3 \\
				
				& PRIMA            
				& \win{}85.415 & \win{}54.911 & \win{}59.479 & \win{}70.038 & \win{}66.356 & \textbf{56.371} & \win{}82.483 & \win{}82.064 & \win{}58.559 & -- & -- & -- & -- & -- & -- & -- & -- & -- & 8/1/0 & 7.7\% & 5.8 \\
				
				& DATIS$_{\text{r}}$ 
				& \win{}81.452 & \win{}53.839 & \win{}58.758 & \win{}75.743 & \win{}69.980 & \win{}54.923 & \win{}81.832 & \win{}80.138 & \win{}47.669 & \win{}85.933 & \win{}85.161 & \win{}83.511 & \win{}81.249 & \win{}66.645 & \win{}73.977 & \textbf{95.728} & \win{}77.686 & \win{}83.176 & 17/0/1 & 7.2\% & 7.9 \\

				& \textbf{\textsc{DuFP}}    
				& \textbf{87.468} & 55.919 & \textbf{66.774} & \textbf{78.830} & \textbf{72.466} & 55.559 & \textbf{90.870} & \textbf{90.036} & 65.841 & \textbf{87.368} & \textbf{86.644} & \textbf{85.072} & \textbf{86.337} & \textbf{74.818} & \textbf{76.454} & 95.595 & \textbf{81.771} & \textbf{84.714} & -- & -- & 1.2 \\
				\bottomrule
			\end{tabular}
			
			\begin{tablenotes}
				\item [1.] 
				For each setting, the best APFD performance is marked in \textbf{bold}, while baseline methods significantly outperformed by \textsc{DuFP} ($\delta \geq 0.147$ and $p\text{-}value \leq 0.05$) are marked with a light grey \tightcolorbox{grey}{blocks\hspace{1pt}}.
				
				\item [2.] 
				The W/T/L denotes the number of cases where the APFD of \textsc{DuFP} is significantly higher, approximately equivalent, or significantly lower than the corresponding baseline, respectively.
				
				\item [3.] 
				The Imp. and Rank columns on the right show the average percentage improvement of \textsc{DuFP} over each method and the average rank of each method across all experimental settings, respectively.
			\end{tablenotes}
		\end{threeparttable}
	}
\end{table*}

To mitigate the effect of randomness, each experimental setting was repeated five times, and the average APFD performance was reported in Table~\ref{tab_apfd}.
The best results are highlighted in bold. 
A grey background indicates that \textsc{DuFP} achieves a large ($\delta \geq 0.147$) and statistically significant ($p\text{-}value \leq 0.05$) improvement over the corresponding baseline.
Across the 18 prioritization settings examined, \textsc{DuFP} achieves the best APFD in 14 cases and the second-best in 4.
Overall, \textsc{DuFP} attains an average rank of 1.2, substantially outperforming the second-ranked method at 5.8.

Among these approaches, the coverage-based methods exhibit the weakest performance.
This is consistent with previous findings that structural coverage has limited capability for fault detection. 
Similarly, although surprise-based prioritization shows improvement over coverage-based methods, its performance remains relatively limited. 
For both the standard SA metrics and the enhanced per-class variants, the APFD results are generally below 70.
Nevertheless, in the adversarial scenario, coverage demonstrates unexpectedly competitive performance. 
For instance, in the adversarial settings of SVHN and Fashion, NAC-CTM obtains the highest and second-highest APFD, respectively. 
This may be attributed to adversarial perturbations incidentally activating rare internal states that are captured by coverage metrics.

Unexpectedly, recent state-of-the-art methods fail to demonstrate consistent advantages over classical uncertainty metrics, despite generally demanding additional information and computational resources.
On average, uncertainty-based methods achieve overall rankings of 5.8 to 8.3, outperforming state-of-the-art baselines designed specifically for prioritization, which rank between 5.8 and 10.7.
Nevertheless, RTS, PRIMA, and the neighborhood-based method DATIS$_{\text{r}}$ each achieve the best result in one setting.
This indicates that uncertainty-based methods offer more stable performance overall, while state-of-the-art methods can achieve strong performance in individual settings.
These findings suggest that existing prioritization approaches remain limited in achieving both effective fault-inducing pattern capture and stable ranking performance.

In contrast, \textsc{DuFP} consistently achieves top-tier performance across all evaluated settings and attains statistical optimality in the majority of cases.
In terms of average performance, our approach achieves at least 42.0\% and 48.3\% APFD improvements over coverage-based and surprise-based methods, respectively. 
Compared with uncertainty-based methods, \textsc{DuFP} achieves an average improvement of 4.6\%--7.9\%, and compared with state-of-the-art mechanism based methods, it achieves an average improvement of 7.2\%--32.3\%. 
Across different data scenarios, methods typically attain higher APFD values on clean data than on shifted data.
This occurs because corrupted and adversarial data reflect distribution shifts in real-world settings, which inevitably reduce the model's capacity to capture feature patterns. 
Even so, driven by the complementary nature of atypicality to ambiguity, \textsc{DuFP} maintains competitive performance under distribution shift conditions.
Furthermore, \textsc{DuFP} achieves superior performance on both image and text datasets, confirming its robust generalization across diverse modalities and data scenarios.
In summary, these results demonstrate that \textsc{DuFP} outperforms state-of-the-art test input prioritization techniques for DNNs.

\begin{tcolorbox}[left=1mm,right=1mm,top=1mm,bottom=1mm]  
	\textbf{Answer to RQ1:} \textsc{DuFP} significantly outperforms baseline methods in prioritizing DNN test inputs and maintains robust generalization across diverse modalities and data scenarios.
\end{tcolorbox}

\subsection{RQ2: Selection}
To assess the effectiveness of \textsc{DuFP} under constrained testing budgets, a comparative evaluation was conducted on the test input selection task. 
The testing budgets were set to 1\%, 3\%, 5\%, 10\%, and then increased in steps of 10\% up to 90\% of the entire candidate set. 
For methods originally designed for prioritization, selection results were derived from the top-ranked inputs produced by the prioritization process.

\begin{figure}[h]
	\centering
	\includegraphics[width=0.95\linewidth, clip, trim=0 0 0 0 ]{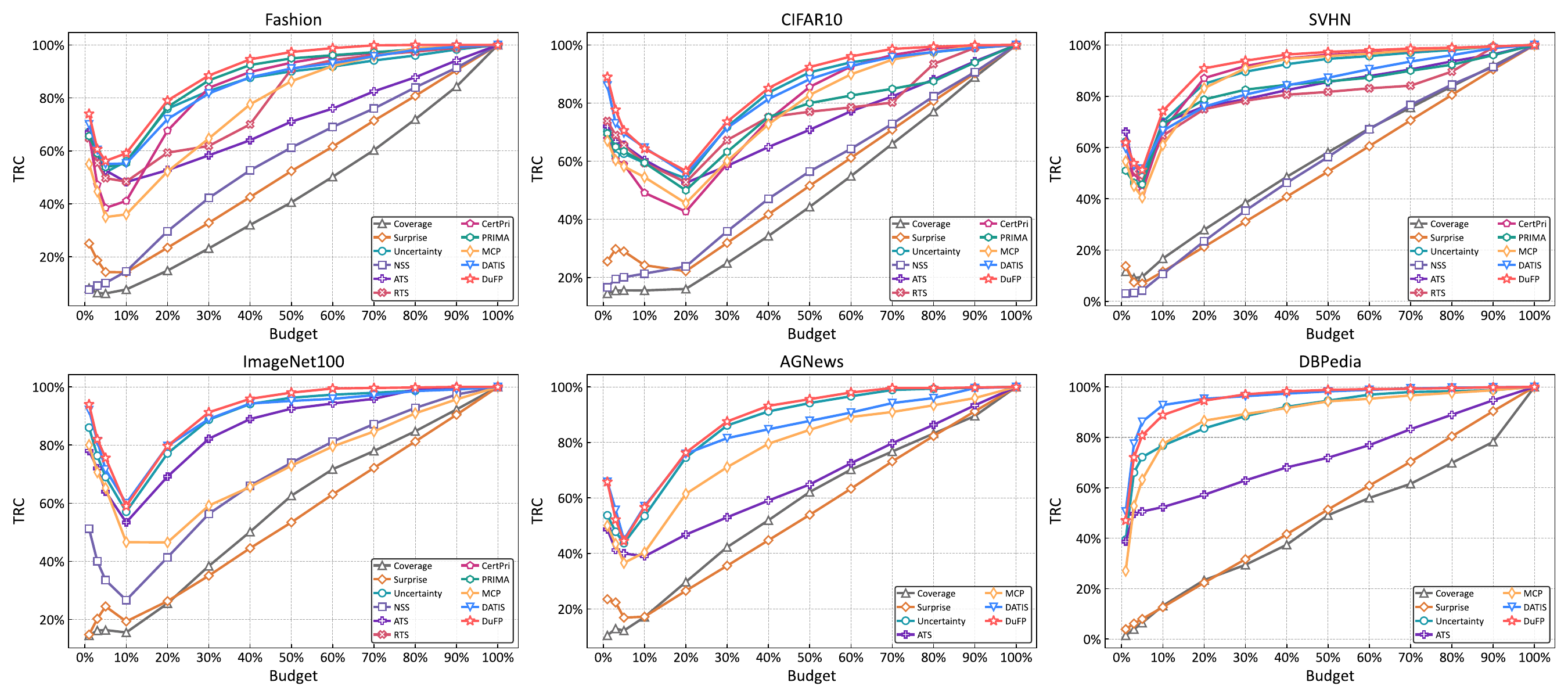}
	
	\caption{The TRC of Test Input Selection under Different Budgets.}
	\label{fig_trc}
	\Description{None.}
\end{figure}

Figure \ref{fig_trc} presents the TRC performance under the nominal setting. 
Results under other settings are available on our project homepage.
The horizontal axis represents the testing budget, and the vertical axis represents the corresponding TRC values. 
For clarity, the coverage, surprise, and uncertainty curves represent the average performance of their respective categories, while the state-of-the-art prioritization approaches (ATS, RTS, NSS, CertPri, PRIMA) and selection-specific methods (MCP, DATIS) are plotted individually.
Overall, \textsc{DuFP} maintains superior performance in the selection task. 
It can be observed that the red curve representing \textsc{DuFP} remains higher than the other curves in most cases. 
This demonstrates that our approach can detect more misclassified instances compared with the other methods under the same budget. 
In other words, \textsc{DuFP} identifies more fault instances at a lower testing cost. 
Nevertheless, it still faces challenges in certain cases. 
On the DBPedia dataset, the TRC of \textsc{DuFP} is surpassed by DATIS. 
Even so, our approach continues to achieve the second-best performance on this dataset. 
The observed performance is similar to the APFD results reported in RQ1, suggesting that APFD and TRC are positively correlated to some extent. 
Both metrics show a preference for prioritizing fault-inducing inputs. 
Notably, TRC is a relative metric that quantifies the maximum fraction of faults detected from the entire fault set or the selected subset.
The TRC values of most methods decline initially and rise subsequently. This indicates that with a smaller selection budget, the proportion of fault samples is higher. 
With increasing budget, the proportion of test inputs exhibiting distinct fault-prone characteristics declines, and the corresponding TRC tends to stabilize.

In summary, these results confirm that \textsc{DuFP} achieves outstanding selection performance across different datasets and budgets.
\begin{tcolorbox}[left=1mm,right=1mm,top=1mm,bottom=1mm]  
	\textbf{Answer to RQ2:} \textsc{DuFP} exhibits superior performance in test input selection. Under the same budget, it generally detects more faults than baseline methods.
\end{tcolorbox}

\subsection{RQ3: Efficiency}

The execution time of \textsc{DuFP} and all baseline methods was measured to assess efficiency. 
The measured time covers the entire prioritization procedure for the candidate set.

\begin{table}[t]
	\centering
	\caption{Time Cost of Test Input Prioritization}
	\label{tab_efficiency}
		\resizebox{\columnwidth}{!}{ 
			\begin{threeparttable}
				\setlength{\tabcolsep}{2pt}   
				\renewcommand{\arraystretch}{1.1}
				{\scriptsize
					\begin{tabular}{c c *{6}{c}}
						\toprule
						\multicolumn{2}{c}{\textbf{Method}} & Fashion & CIFAR\mbox{-}10 & SVHN & ImageNet\mbox{-}100 & AGNews & DBPedia \\
						\midrule
						
						\multirow{4}{*}{\rotatebox{90}{Coverage}}
						& NAC\mbox{-}CTM   & 8.19\,s & 4.42\,s & 1\,m\,3\,s & 24.15\,s & 8.93\,s & 1\,m\,21\,s \\
						& NAC\mbox{-}CAM   & 7.04\,s & 7.92\,s & 13\,m\,39\,s & 22.38\,s & 9.62\,s & 1\,m\,30\,s \\
						& KMNC\mbox{-}CTM  & 11\,m\,47\,s & 12\,m\,9\,s & 3\,h\,49\,m & 5\,m\,55\,s & 11\,m & 1\,h\,25\,m \\
						& KMNC\mbox{-}CAM  & 2\,h\,18\,m & 2\,h\,6\,m & >100\,h & >10\,h & 1\,h\,7\,m & >10\,h \\
						\midrule
						
						\multirow{6}{*}{\rotatebox{90}{Surprise}}
						& LSA              & 59.61\,s & 52.05\,s & 26\,m\,27\,s & 6\,m\,55\,s & 21\,m\,43\,s & >10\,h \\
						& PC\mbox{-}LSA    & 17.26\,s & 14.65\,s & 3\,m\,47\,s & 5\,m\,42\,s & 6\,m\,46\,s & 1\,h\,3\,m \\
						& PC\mbox{-}MLSA   & 16.10\,s & 13.83\,s & 2\,m\,26\,s & 5\,m\,42\,s & 6\,m\,21\,s & 33\,m\,2\,s \\
						& PC\mbox{-}DSA    & 23\,m\,23\,s & 17\,m\,39\,s & 9\,h\,52\,m & 7\,m\,40\,s & 35\,m\,43\,s & >10\,h \\
						& PC\mbox{-}MDSA   & 31.99\,s & 31.94\,s & 25\,m\,49\,s & 5\,m\,27\,s & 2\,m\,28\,s & 13\,m\,16\,s \\
						& PC\mbox{-}MMDSA  & 46.02\,s & 44.66\,s & 29\,m\,8\,s & 5\,m\,54\,s & 3\,m\,9\,s & 15\,m\,56\,s \\
						\midrule
						
						\multirow{6}{*}{\rotatebox{90}{Uncertainty}}
						& DeepGini         & 0.03\,s & 0.03\,s & 0.06\,s & 0.02\,s & 0.01\,s & 0.06\,s \\
						& MaxP             & 0.03\,s & 0.02\,s & 0.06\,s & 0.01\,s & 0.01\,s & 0.06\,s \\
						& Margin           & 0.03\,s & 0.03\,s & 0.06\,s & 0.02\,s & 0.01\,s & 0.07\,s \\
						& Entropy          & 0.03\,s & 0.03\,s & 0.06\,s & 0.02\,s & 0.01\,s & 0.06\,s \\
						& FAST             & 7.10\,s & 6.83\,s & 15\,m\,24\,s & 46\,m\,55\,s & 16\,m\,55\,s & 1\,h\,32\,m \\
						& NNS              & 4.63\,s & 4.05\,s & 12.4\,s & 1.03\,s & 1.02\,s & 10.76\,s \\
						\midrule
						
						\multirow{7}{*}{\rotatebox{90}{Mechanism}}
						& NSS              & 4.86\,s & 5.09\,s & 12.4\,s & 1\,m\,54\,s & -- & -- \\
						& ATS              & 9.44\,s & 10.50\,s & 1\,m\,29\,s & 43\,m\,3\,s & 0.17\,s & 15.24\,s \\
						& RTS              & 10\,m\,12\,s & 11\,m\,25\,s & 36\,m\,42\,s & -- & -- & -- \\
						& CertPri          & 15\,m\,6\,s & 15\,m\,37\,s & 53\,m\,3\,s & -- & -- & -- \\

						& PRIMA            & 31\,m\,54\,s & 48\,m\,23\,s & 1\,h\,11\,m & -- & -- & -- \\
						& DATIS$_{\text{r}}$ & 17.85\,s & 14.62\,s & 50.87\,s & 5\,m\,53\,s & 1\,m\,58\,s & 24\,m\,51\,s \\
					
						
						& \textbf{\textsc{DuFP}}    & \textbf{12.82\,s} & \textbf{10.77\,s} & \textbf{35.48\,s} & \textbf{4\,m\,54\,s} & \textbf{1\,m\,53\,s} & \textbf{19\,m\,5\,s} \\
						\bottomrule
					\end{tabular}
				}
			\end{threeparttable}
		}
	\end{table}

Table~\ref{tab_efficiency} reports the time cost of each approach, with that of \textsc{DuFP} highlighted in bold.
In general, prioritization techniques focus on testing well-trained DNNs, with most remaining independent of the training process. 
As a result, the time overhead introduced is relatively modest.

The time cost of coverage-based methods is influenced by both the coverage metric and the prioritization strategy. 
Among them, NAC achieves lower cost due to its coarse granularity, whereas KMNC with finer granularity incurs significantly higher computational overhead. 
Under the same coverage metric, the CAM strategy typically results in higher time cost than CTM due to its dynamic prioritization.
Moreover, DNNs with more complex architectures require the extraction of a broader activation space, leading to higher computational overhead.
Similarly, surprise based methods that operate on single-layer activations (e.g., PC-MLSA) achieve substantially lower time cost than those utilizing all intermediate activations (e.g., PC-DSA).

Uncertainty-based methods are the most efficient among all approaches. Classical metrics such as DeepGini and Entropy compute directly from prediction probabilities, enabling prioritization within 0.01\,s to 0.07\,s. 
Calibration-based methods require more time: NNS generally takes under 15\,s with minimal sensitivity to the model architecture, whereas FAST ranges from 6.83\,s on CIFAR10-ResNet20 to 1\,h\,32\,m on DBPedia-BERT depending on the architecture and the chosen layer.

For mechanism-based methods, ATS and NSS are relatively efficient, whereas RTS, CertPri, and PRIMA are considerably more time-consuming, often exceeding 10\,m and sometimes exceeding 1\,h.
This overhead arises because RTS requires computing the similarity between each test input and the training set, CertPri applies adversarial perturbations to each input individually, and PRIMA generates a large number of mutants for both inputs and models.

In contrast, \textsc{DuFP} demonstrates substantially higher efficiency with limited sensitivity to model complexity, as its computations are confined to a fixed-dimensional feature space.
\textsc{DuFP} completes prioritization within 40\,s on small-scale image datasets and takes from 1\,m\,53\,s to approximately 20\,m on complex transformer based models.
The time cost is primarily dominated by feature extraction and KNN search, both of which depend on the training set size.
For example, the only case exceeding 5\,m is DBPedia (19\,m\,5\,s), where the large number of training and test samples increases the KNN computation cost.
Moreover, the forward inference and feature extraction over the training set constitute a one-time offline stage, whose results can be precomputed and reused across candidate sets in practical deployment.
On larger benchmarks such as ImageNet-100, this offline stage dominates the reported cost, and the limited online overhead remains acceptable given the stable improvements over uncertainty baselines reported in Table~\ref{tab_apfd}.
Overall, prioritization with \textsc{DuFP} is accomplished within 5\,m on five of the six datasets, demonstrating superior efficiency over most baseline methods.

\begin{tcolorbox}[left=0.5mm,right=0.5mm,top=0.5mm,bottom=0mm]  
	\textbf{Answer to RQ3:} \textsc{DuFP} achieves efficient test input prioritization with an acceptable time cost and proves faster than numerous state-of-the-art approaches.
\end{tcolorbox}

\subsection{RQ4: Ablation Study}

\begin{figure}[h]
	\centering
	\includegraphics[width=0.95\linewidth, clip, trim=0 0 0 0 ]{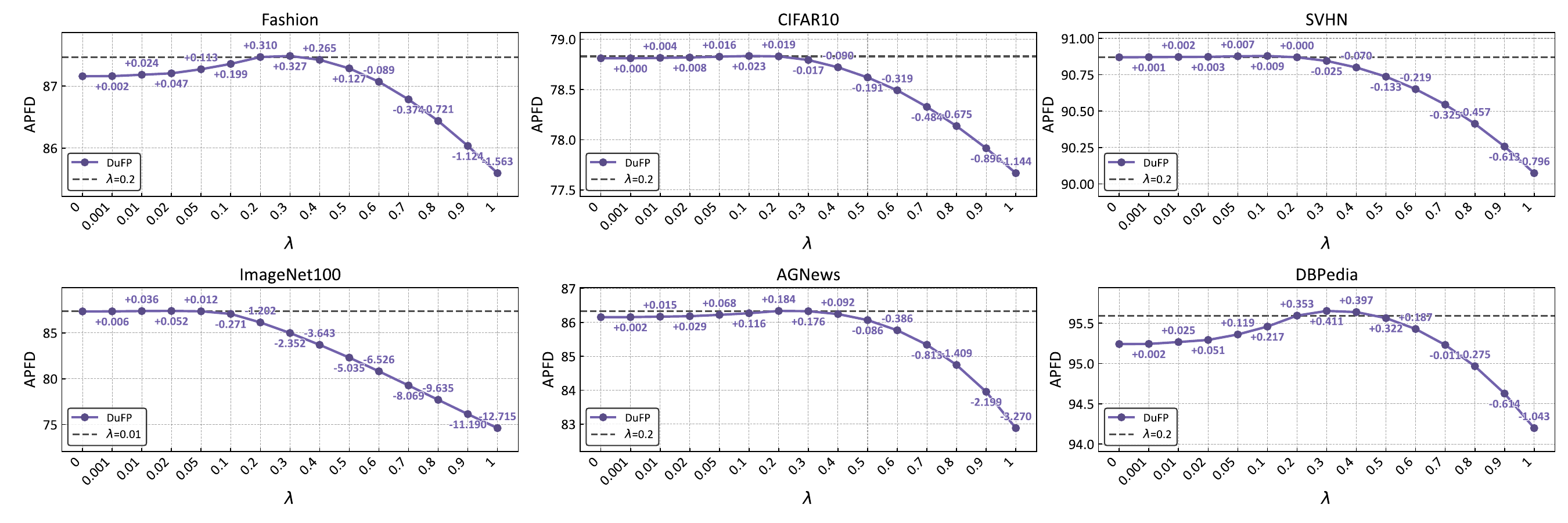}
	\caption{APFD Performance under Different $\lambda$ Values.}
	\label{fig_lambda}
	\Description{None.}
\end{figure}

\begin{figure}[h]
	\centering
	\includegraphics[width=0.95\linewidth, clip, trim=0 0 0 0 ]{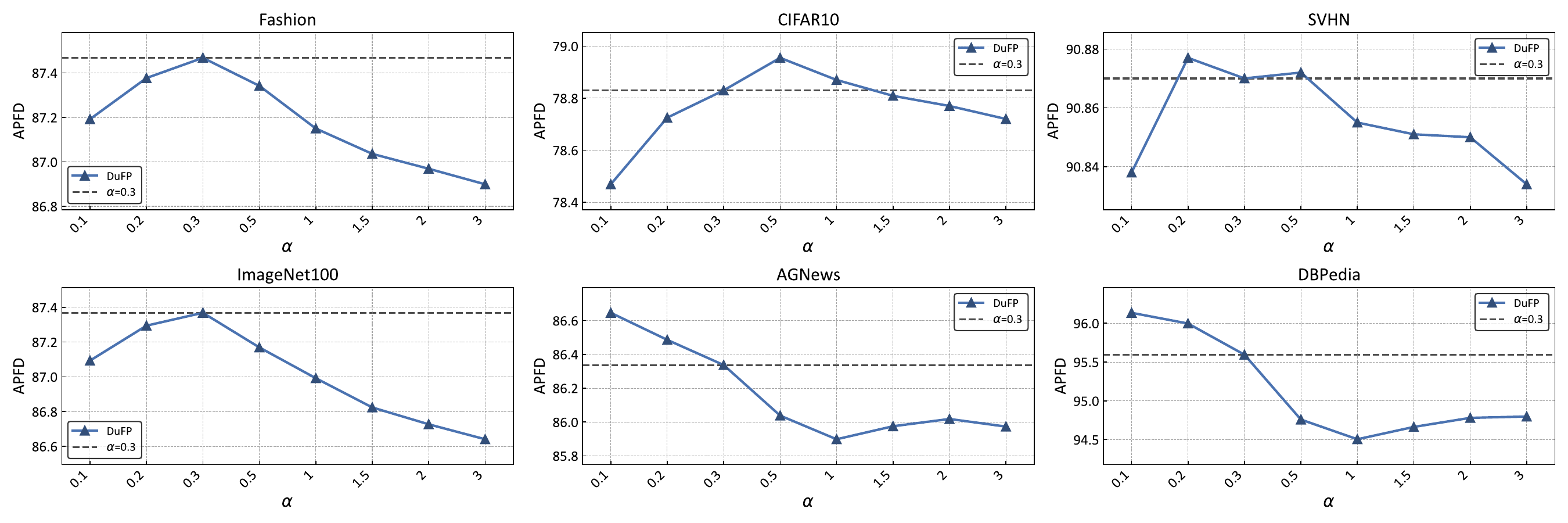}
	\caption{APFD Performance under Different $\alpha$ Values.}
	\label{fig_alpha}
	\Description{None.}
\end{figure}

\begin{figure*}[h]
	\centering
	\includegraphics[width=0.95\linewidth, clip, trim=10 10 10 10]{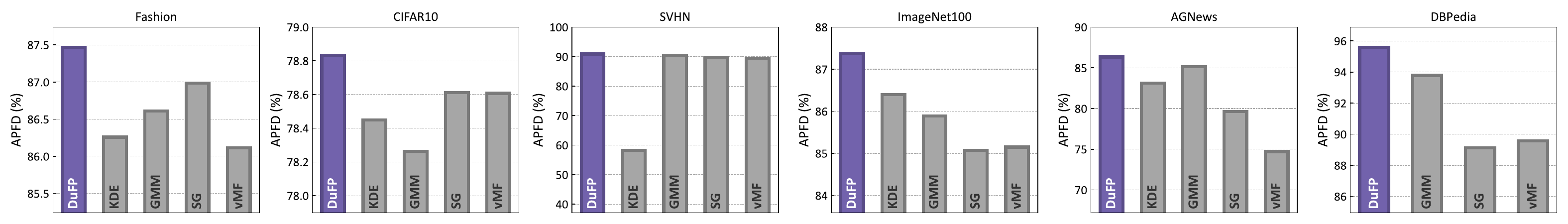}
	
	\caption{APFD performance of \textsc{DuFP} with different density estimation methods.}
	\label{fig_density}
	\Description{None.}
\end{figure*}

\subsubsection{Effect of $\lambda$.}
The hyperparameter $\lambda$ controls the balance between the two components of the hybrid uncertainty.
A smaller $\lambda$ biases the hybrid uncertainty toward decision ambiguity, while a larger $\lambda$ emphasizes distributional atypicality.
In this experiment, \textsc{DuFP} was evaluated with $\lambda$ values ranging from 0 to 1 in increments of 0.1, supplemented by finer-grained values of 0.001, 0.01, 0.02, and 0.05.

Figure \ref{fig_lambda} presents the APFD performance of \textsc{DuFP} across different $\lambda$ values, with the improvement relative to $\lambda = 0$ annotated beside each data point.
The APFD curves exhibit unimodal patterns, characterized by an initial increase followed by a gradual decrease. 
This indicates that the best APFD is typically achieved at intermediate $\lambda$ values, rather than at the extreme values of $\lambda = 0$ (ambiguity only) or $\lambda = 1$ (atypicality only).
This demonstrates that the effectiveness of \textsc{DuFP} arises from the combined contributions of both perspectives, where ambiguity provides the primary signal and atypicality serves as a complementary cue that rectifies the blind spots of ambiguity.
The default $\lambda = 0.2$ was selected via held-out validation over the same grid of values, with 10\% of the training data reserved as a labeled validation set.
Owing to the robustness of \textsc{DuFP} to $\lambda$, this value is adopted as a universal empirical value across benchmarks without per-subject re-validation.
An exception is ImageNet-100, where the same procedure selected $\lambda = 0.01$, as its high feature dimensionality together with numerous classes and scarce per-class samples yields a sparser per-class feature space.
For new subjects of the same modality with conventional class scales, we recommend directly reusing these empirical values.
For regimes with high class counts and sparse per-class samples or for unexplored modalities, re-selecting $\lambda$ through the same validation procedure is suggested.


\subsubsection{Effect of $\alpha$.}

In KNN density estimation, fixing a global neighborhood ratio $\eta$ inevitably leads to sub-optimal local density estimates across datasets of varying scales.
As demonstrated by \cite{cover1967nearest}, the neighbor count should scale sub-linearly with the data volume to ensure statistical stability.
Following the widely adopted square-root scaling heuristic \cite{hart2001pattern}, a universal scaling constant $\alpha$ is introduced to formulate the adaptive scaling law for the ratio hyperparameter: $\eta = {\alpha}/{\sqrt{N}}$.
In this experiment, $\alpha$ was evaluated at values of 0.1, 0.2, 0.3, 0.5, 1, 1.5, 2, and 3.

Figure \ref{fig_alpha} presents the APFD performance of \textsc{DuFP} across different $\alpha$ values.
Overall, the performance fluctuation remains within 2\%, indicating that \textsc{DuFP} is insensitive to this parameter.
Among the tested values, a relatively small positive value of $\alpha$ tends to yield strong performance, suggesting that the local neighborhood distribution is more informative for density estimation.
The default $\alpha = 0.3$ was determined through a similar hyperparameter selection procedure and adopted as a universal empirical value across all datasets, consistent with the superior region observed in Figure \ref{fig_alpha}.


\subsubsection{Density Estimation Method.}
Density estimation is a fundamental component of \textsc{DuFP}. 
To evaluate the impact of the density estimation method, the KNN estimator in \textsc{DuFP} was replaced with four representative alternatives, including Kernel Density Estimation (KDE), Single Gaussian (SG), Gaussian Mixture Model (GMM), and the von Mises-Fisher (vMF) distribution \cite{vmf_density}.
These models cover parametric, mixture-based, directional, and nonparametric density families. 
As shown in Figure \ref{fig_density}, \textsc{DuFP} with its default KNN estimator consistently achieves the highest APFD across all settings. 
This confirms that the KNN-based density estimator effectively captures neighborhood information in the feature space.

\begin{tcolorbox}[left=1mm,right=1mm,top=1mm,bottom=1mm]  
	\textbf{Answer to RQ4:}
	Each design element of \textsc{DuFP} contributes to its overall effectiveness.
	The adopted configuration achieves robust performance across the majority of evaluation settings.
\end{tcolorbox}

\section{Discussion}
\label{sec_discussion}

\subsection{Class Imbalance Analysis}
\label{sec_discussion_imbalance}
Since \textsc{DuFP} relies on class-conditional density estimation, its behavior under class-imbalanced training data merits discussion.
The adaptive neighbor count $k_c = n_c \cdot \eta$ introduced in Section~\ref{sec_method_density} removes the explicit class-size dependence from \eqref{eq_density}.
Density estimates therefore reflect geometric proximity rather than class size and remain comparable between majority and minority classes.
Substituting \eqref{eq_density} into the posterior ratio \eqref{eq_ambiguity} makes this property explicit:
\begin{equation}\label{eq_amb_decompose}
	\log s_{\mathrm{amb}}(z) = \underbrace{d \log \frac{r_{\hat c}(z)}{r_{c^*}(z)}}_{\text{geometric term}} + \underbrace{\log \frac{n_{c^*}}{n_{\hat c}}}_{\text{prior correction}}
\end{equation}
where the geometric term is independent of class frequency and the prior term corresponds to the standard Bayes correction for class imbalance.
The atypicality score $s_{\mathrm{atyp}}(z) = V_d\, r_{\hat c}(z)^d / \eta$ likewise depends only on the geometric support of the predicted class.
Class prevalence therefore influences \textsc{DuFP} only through the prior term.
\textsc{DuFP} is thus expected to remain effective under long-tailed distributions, although density estimates of extremely scarce classes exhibit higher variance as fewer neighbors are available.

\subsection{Threats to Validity}
The internal validity mainly lies in the implementation of \textsc{DuFP} and the baseline methods. 
To mitigate this threat, we carefully verified the implementation of \textsc{DuFP} to ensure correctness. 
For the baselines, most implementations were obtained from their official repositories, and those without publicly available code were re-implemented following the original papers.
All source code was checked against the original descriptions to ensure faithfulness.
The hyperparameters were configured following the settings reported in previous studies. 
Moreover, identical experimental settings were maintained across all methods to ensure a fair comparison.
The external validity mainly lies in the DNN models and datasets adopted in the evaluation. 
To mitigate this threat, the evaluation was conducted on {six} representative model architectures across image and text datasets. 
For each dataset, the proposed method was evaluated on nominal, corrupted, and adversarial candidate sets, covering both clean and distribution shift conditions.
Finally, the conclusions were supported by comparisons with {23} representative peer techniques.

\subsection{Limitations and Future Work}
Experimental results demonstrate that \textsc{DuFP} achieves superior performance in prioritizing test inputs for DNNs. 
However, we recognize that \textsc{DuFP} still has a few limitations that could be further improved.
In terms of effectiveness, \textsc{DuFP} falls short of the best performance in some evaluation settings. 
Future work will investigate weighted density estimation based on prediction confidence, which may mitigate the impact of outliers and enhance robustness. 
In terms of efficiency, the runtime of \textsc{DuFP} stays within 20 minutes, yet it remains longer than naïve uncertainty methods, which typically require less than one second. 
To reduce the time cost, density estimation on a representative subset of the training data could be explored.
In terms of scalability, \textsc{DuFP} demonstrates applicability to image and text datasets but is not evaluated on other data modalities. 
Future research will extend the evaluation to other modalities such as speech and graphs.
In terms of task scope, \textsc{DuFP} targets closed-set classification, where a fault is defined as a misclassification and a prediction takes the form of a discrete class.
\textsc{DuFP} therefore cannot be directly applied to regression or generative systems, because their faults are no longer misclassifications and the predicted class and class-conditional densities on which it operates are undefined.
Nevertheless, the core idea of capturing ambiguity and atypicality within the feature neighborhood generalizes beyond classification, since both signals stem from the supervisory information carried by neighboring training samples rather than from discrete labels themselves.

\section{Conclusion}
\label{sec_conclusion}
In this paper, we propose \textsc{DuFP}, an effective test input prioritization method for DNNs from dual perspectives.
\textsc{DuFP} captures fault-inducing characteristics through class-wise density estimation built upon model information and supervisory signals from the training data.
Based on the density estimation, both decision ambiguity and distributional atypicality are quantified.
The resulting hybrid uncertainty enables robust assessment of prediction correctness for test inputs.
We conducted comprehensive evaluations on image and text datasets across clean, corrupted, and adversarial settings. 
Experimental results demonstrate that \textsc{DuFP} outperforms state-of-the-art approaches while maintaining relatively low time overhead. 
We believe that \textsc{DuFP} enhances the efficiency of DNN testing and contributes to improving the reliability of DNN-based systems.

\section{Data Availability Statement}
To facilitate reproducibility, we publicly release the source code and all data resources used in this study, including datasets and DNN models.
The replication package is hosted at Zenodo \cite{DuFP_artifacts} with doi \url{10.5281/zenodo.19184936}.


\makeatletter
\let\ase@original@lbibitem\@lbibitem
\def\@lbibitem[#1]#2{%
  \ase@original@lbibitem[#1]{#2}%
  \def\ase@currentbibkey{#2}%
  \def\ase@balancebibkey{pgd}%
  \ifx\ase@currentbibkey\ase@balancebibkey
    \balance
  \fi
}
\makeatother
\bibliographystyle{ACM-Reference-Format}
\bibliography{reference}




\end{document}